\documentclass[12pt]{article}

\usepackage{amsmath,amssymb}

\title{
Coherent mixing in  three and four quark  generations }
\author{Yu.A.Simonov\\ Institute of Theoretical and Experimental
Physics\\ 117118, Moscow, B.Cheremushkinskaya 25, Russia}
\date{}

\newcommand{\beq}{\begin{eqnarray}}
 \newcommand{\eeq}{\end{eqnarray}}
\newcommand{\be}{\begin{equation}}
 \newcommand{\ee}{\end{equation}}

 \newcommand{\la}{\lesssim}
 \newcommand{\ga}{\gtrsim}

\newcommand{{\SD}}{\rm SD}

\newcommand{\lan}{\langle}
\newcommand{\ran}{\rangle}

\begin{document}
\maketitle
\begin{abstract}
New  dynamical mechanism of quark mass generations and mixing is demonstrated
in the examples of three and four generations. In the framework of the   new
mixing pattern, called the coherent mixing, the CKM elements are predicted
compatible with experimental data for three  generations, and are strongly
constrained for four generations.

\end{abstract}

1. The idea that quark masses can be obtained from composite
scalar fields instead of  elementary Higgs fields is actively
developed last decades, for reviews and references see \cite{1,2}.

The basic underlying mechanism for the spontaneous electroweak
symmetry breaking (EWSB) and mass generation was the effective
quartic fermion interaction which is bosonized in the standard
way. What is left unanswered in this approach is the  dynamical
origin of generations and  the peculiar pattern of masses and
mixings in  the observed up to now  three quark generations.

It is a purpose of the present letter to describe, basing on the
recent paper [3], the mechanism which produces three or more
generations of quarks and can predict intervals  of values for
mixing coefficients and masses of $t',b'$ in the case of the
fourth generation.

One starts with the unbroken gauge field  $C_\mu$  (it can be taken as $SU(N)$
field, but this is not important for  what follows) and quark Lagrangian
$L=g\bar \psi_a \gamma_\mu C^{ab}_\mu \psi_b$, where fields are organized into
two sectors $\psi_a^{(i)}, i=1,2.$

 \be
 \psi_{Li}^{(1)} = \left( \begin{array}{l}u_L\\d_L\end{array}\right), \left( \begin{array}{l}c_L\\s_L\end{array}\right),
 \left( \begin{array}{l}t_L\\b_L\end{array}\right); \psi_R^{(1)} = (d_R, s_R,
 b_R)\label{1}\ee $$
\psi_{Li}^{(2)} = \left( \begin{array}{l}d^c_L\\u^c_L\end{array}\right), \left(
\begin{array}{l}s^c_L\\c^c_L\end{array}\right),
 \left( \begin{array}{l}b^c_L\\t^c_L\end{array}\right); \psi_R^{(2)} = (u^c_R, c^c_R,
 t^c_R)$$
 where $\psi^c$ means the charge conjugated quark field.

 We assume  that nonperturbative vacuum fields $C_\mu$ can  form connected
 field correlators, such that (vacuum averaging of Euclidean
 fields is implied)
 \be tr\ll C_{\mu_1} (x_1) C_{\mu_2}(x_2) ... C_{\mu_n}(x_n) \gg \equiv
 J_{\mu_1,...\mu_n}^{(n)} (x_1, x_2,... x_n).\label{2}\ee
 We note, that the form (\ref{2}) is in general not gauge invariant, but for
 nonconfining field $C_\mu(x)$ one can define (\ref{2}) in such a way, that in
 the local limit $|x_i -x_j|\to 0, i,j,=1,... n$ one can isolate in (\ref{2}) a
 gauge invariant piece of the quark self-energy (this implies that the scale
 $M$ of $J^{(n)}$ can be taken much larger than any other  mass in the
 problem). As a result one obtains in the partition  function $Z=\int DC D\psi
 D\bar\psi e^L$; the effective quark Lagrangian

 \be \lan \exp \int L_{} d^4x\ran  =\exp \left\{ \sum_{n=2,4,6,...} J_n (x_1,... x_n)
\Psi(x_1)... \Psi(x_n) dx_1... dx_n\right\}.\label{3}\ee

Here $\Psi(x) ={\bar \psi}(x) \gamma_\mu   \psi(x)$, and we have
suppressed spinor and group indices.

Now  making pairwise Fierz transformations and keeping only scalar
and pseudoscalar terms, one can express all quark bilinears in
terms of $\Phi_{RL}$ and $\Phi_{LR}$, where $\Phi_{RL} (x_1, x_2)=
{ \bar {\psi}^a_{R}}(x)  \psi_L^a (x_2)$, superscript $a$ denotes
$SU(2)$ index of EW group or its GUT equivalent. Next one does a
bosonization trick introducing auxiliary functions $\mu(x_i, x_k),
\varphi(x_i, x_k)$ in  $\delta$-function terms and finding the
stationary points of (\ref{3}) in $\mu$ and $\varphi$

$$\lan \exp \int L d^4 x\ran_C= \int
D\varphi D\varphi^+ \tilde \delta (\varphi-\Phi_{RL}) \tilde \delta
(\varphi^+-\Phi_{LR})\times $$ \be\times \exp \left\{-\sum_{n=1,4} \int J_n
(\varphi\varphi^+ + \varphi^+\varphi)^{n/2} dx_1 ... dx_n\right\} ,\label{4}\ee

where  $\tilde \delta (f-F)=  \int D\mu \exp  \left\{ -i \int \mu (f-F)
 dx\right\}$.

  The
final equation can be written in the momentum space for the quark mass function
$\mu(p)$ in the form (see \cite{3} for derivation and more discussion) \be
\mu(p) = \sum_n \int D_n \{p_1, ...p_{n-1}\} \bar J_n \{p; p_1,...
p_{n-1}\}.\label{5}\ee Here $D_n\{...\}$ denotes product $\prod^{n-1}_{k=1}
\frac{d^4p_k}{(2\pi)^4}d(p_k)$,  $d(p) = \frac{\mu(p)}{p^2+\mu^2(p)}$. Note,
that $\mu(p)$ plays the role of the $p$-dependent mass of quark, since it
enters the quark Green's function as $S^{-1} (p) = \hat p + i\mu(p)$.

It  is important, that the stationary points of the wave
functional (\ref{4}) can be written for the real quantity $\hat
\varphi (x_i, x_k)\equiv -i \varphi (x_i, x_k)$ in the local limit
$(M\to \infty)$ as follows

\be(-^{\framebox{}}~+\mu^2)\widehat{\varphi} =- \frac{\delta
U\{\widehat{\varphi}\}}{\delta \widehat{\varphi} (x)};~
U\{\widehat{\varphi}\}= - \frac12 \tilde J_2 \widehat{\varphi}^2+
\frac14 \tilde J_4 \widehat{\varphi}^4-\frac16  \tilde J_6
\widehat{\varphi}^6+...\label{6}\ee

At this point one realizes that $\tilde J_n$ are positive for
connected correlators and therefore $U\{\hat \varphi\}$ can have
multiple minima $\hat \varphi_i$. We can associate composite
scalars of generations with these minima, and then solutions of
(\ref{6}) of kink type, which connect different minima $i,k$ of
$U\{\hat \varphi\}$ can be related to mixing solutions of
composite scalars $\hat \varphi_{ik}$. The corresponding masses
$\mu_{ii}\equiv \mu_i$ and $\mu_{ik}$ can be found from the
equation $\hat \mu = (^{\framebox{}}~+ \hat \mu^2) \hat \varphi$
where both $\hat \mu$ and $\hat \varphi$ are matrices in
generation indices. In this way one obtains the initial mass
matrix $\mu_{ik}$, which should be diagonalized to yield the final
physical mass matrix $\hat m$ and CKM mixing coefficients
$V_{ik}$. We shall not make explicit at this point the
coefficients $\tilde J_n$ and the functional $U\{\hat \varphi\}$,
referring it to later publications, but rather shall try to guess
the form of the matrix $\hat \mu$, corresponding to the realistic
physical masses and CKM coefficients.

Namely,  because of the strong hierarchy of masses, $\mu_{33} \gg \mu_{22}\gg
\mu_{11}$, the matrix solution of (\ref{5})  $\mu_{ik}$ acquires the
approximate form $\mu_{ik} = \sqrt{\mu_i\mu_k}$, with $\mu_i\equiv \mu_{ii}$
and $\mu_k\equiv \mu_{kk}$, and one can write a slightly distorted form \be
\mu_{ik} =\sqrt{\mu_i \mu_k (1+\eta_{ik})} , ~~ i\neq k;
|\eta_{ik}|<1.\label{7}\ee

We shall call (\ref{7}) as in \cite{3} the Coherent Mixing Mechanism (CMM), and
apply it to the case of three and four generations.

2. In the case of three generations one can expand the eigenvalue equation
$\det (\hat \mu -m \hat 1)=0$ to the lowest orders in $\eta_{ik}$ to find the
physical masses $m_i$

 \be m^3 - m^2 \sigma + m\xi -\zeta =0\label{8}\ee
with  $\sigma=\sum^3_{r=1}\mu_i,$ $ \xi = - \sum_{i\neq j} \mu_i \mu_j
\eta_{ij}, $ $ \zeta =\mu_1\mu_2\mu_3\left(-\frac14 \sum_{i\neq j} \eta^2_{ij}+
\frac12 \sum_{i\neq j~ l\neq k} \eta_{ij} \eta_{lk}\right)$.

For  $m_1$  to be positive and for the natural hierarchy
   $m_1 \ll m_2 \ll m_3$, one needs
$\eta_{ik} <0, i,k=1,2,3$ and choosing $\eta_{13} = \eta_{23} \equiv-\eta,~~
\eta_{12} =-\delta, ~~ 0<\delta\ll \eta$, one has \be m_1 \approx \mu_1 \delta,
~~m_2\approx \frac{\mu_3}{m_3} (\mu_1+\mu_2) \eta, ~~ m_3 \approx
\mu_1+\mu_2+\mu_3.\label{9}\ee It is interesting, that for $\delta \ll \eta \ll
1 $ the CMM has made the hierarchy much more pronounced, than original
situation with $\mu_1< \mu_2< \mu_3$, and $m_1, m_2$ can be made very close to
zero,  while $m_3$ is not far from $\mu_3$.

The diagonalizing matrix $W$, defined as $\hat \mu = W^+ \hat m
W$, is given in the appendix 2 of \cite{3} together with the
resulting CKM matrix $V_{CKM} = W_uW^+_d$.

The matrix $W$ can be conveniently written in terms of elements $\mu_{ik}$ and
$m_i$, using the  method of \cite{4},  in CMM this is simplified in the limit
$\delta \to 0$, yielding  condition $W_{31}=0$, and an equivalent limiting form
for the $ 4 \times 4$ matrix $W$ has three zeros: $W_{31} = W_{41} =W_{42} =0$

\be W =\left( \begin{array}{lll} c_\alpha,& - s_\alpha c_\beta
e^{i\delta_{12}},& s_\alpha s_\beta e^{i\delta_{13}}\\
- s_\alpha
e^{-i\delta_{12}},& -c_\alpha c_\beta&, c_\alpha s_\beta e^{i\delta_{23}}\\
0, &s_\beta e^{-i\delta_{23}},& c_\beta\end{array}\right)
\label{10}\ee $W$ is expressed via two sine parameters $s_\alpha,
s_\beta$ and two phases.
 Here the
phases $\delta_{ik}= \delta_i-\delta_k, k=1,2,3$  satisfy conditions

\be \delta_{12}-\delta_{13}+\delta_{23}=0.\label{11}\ee

Parameters $c_\alpha, c_\beta, (s_i=\sqrt{1-c^2_i}, i=\alpha, \beta)$ are
expressed via $m_i, \mu_i, i= 1,2,3$ (similar expressions are found in
\cite{4}).

\be c_\alpha=\sqrt{\frac{m_2-\mu_1}{m_2-m_1}},~~ s_\alpha=\sqrt{1-c^2_\alpha},
c_\beta=\sqrt{\frac{m_3-\mu_2}{2m_3-\mu_2-\mu_3}},~~
s_\beta=\sqrt{1-c^2_\beta}.\label{12}\ee

Surprisingly, the simple form (10) as will be seen yields realistic CKM matrix
for three  generations, which is  written below in two ways: a general form,
and
 another with assumption of $\delta^u_k = \delta^d_k =0$ for $k\neq 1$.

\newcommand{{\sa}}{s_\alpha}

\newcommand{{\sbe}}{s_\beta}
$$V_{ud} = c^u_\alpha c^d_\alpha+c^u_\beta c^d_\beta  s^u_\alpha s^d_\alpha
e^{i\Delta_{12}}+ s^u_\alpha s^u_\beta s^d_\alpha s^d_\beta
e^{i\Delta_{13}}\approx c^u_\alpha  c^d_\alpha +s^u_\alpha s^d_\alpha
e^{i\Delta_{12}}\cos (\beta_u-\beta_d),$$

$$ V_{us} = s_\alpha^uc^u_\beta  c^d_\alpha c^d_\beta e^{i\delta_{12}^u}- c_\alpha^u
s^d_\alpha  e^{i\delta^d_{12}}+s^u_\alpha s^u_\beta c^d_\alpha s^d_\beta
e^{i\delta^u_{13}-i\delta^d_{23}}\approx -c^u_\alpha s^d_\alpha
e^{i\delta^d_{12}} + s^u_\alpha c^d_\alpha e^{i\delta^u_{12}}\cos
(\beta_u-\beta_d)$$

$$V_{ub}= s^u_\alpha s^u_\beta c^d_\beta
e^{i\delta_{13}^u} - s^d_\beta s^u_\alpha c^u_\beta e^{i(\delta_{12}^u+\delta^d_{23})}\approx s^u_\alpha e^{i\delta^u_{13}}
\sin (\beta_u-\beta_d)$$
$$
 V_{cd} =- s^u_\alpha  c^d_\alpha e^{-i\delta^u_{12}}+ c^u_\alpha c^u_\beta
c^d_\beta s^d_\alpha e^{-i\delta^d_{12}}+$$\be+ c^u_\alpha s^u_\beta s^d_\alpha
s^d_\beta e^{i\delta^u_{23}-i\delta^d_{13}}\approx  c^u_\alpha s^d_\alpha
e^{-i\delta^d_{12}}\cos (\beta_u-\beta_d) -s^u_\alpha c^d_\alpha
e^{-i\delta^u_{12}}\label{13}\ee

$$V_{cs}= s^u_\alpha   s^d_\alpha
 e^{-i\Delta_{12}}+ c_\alpha^u c_\beta^u c^d_\alpha
c^d_\beta + c^u_\alpha c^d_\alpha s_\beta^u s^d_\beta e^{i\Delta_{23}}\approx
s^u_\alpha s^d_\alpha e^{-i\Delta_{12}}+ c^u_\alpha c^d_\alpha \cos
(\beta_u-\beta_d),$$

$$V_{cb}=c^u_\alpha c^d_\beta s^u_\beta e^{i\delta^u_{23}}
- c^u_\alpha c^u_\beta  s^d_\beta  e^{i\delta^d_{23}}\approx c^u_\alpha \sin
(\beta_u-\beta_d),$$

$$V_{td}=c^u_\beta s^d_\alpha s^d_\beta  e^{-i\delta_{13}^d} - c^d_\beta s^u_\beta
s^d_\alpha e^{-i\delta^u_{23}-i\delta^d_{12}}\approx s^d_\alpha
e^{-i\delta^d_{13}}\sin (\beta_d-\beta_u),$$

$$V_{ts} = -  s^u_\beta c^d_\alpha
c^d_\beta e^{-i\delta_{23}^u} + c^d_\alpha c^u_\beta s^d_\beta
e^{-i\delta^d_{23}}\approx  c^d_\alpha \sin (\beta_d-\beta_u),$$

$$V_{tb} =
s^u_\beta s^d_\beta e^{-i\Delta_{23}}+ c^u_\beta c^d_\beta\approx \cos
(\beta_u-\beta_d).$$

Here $\Delta_{ij} \equiv \delta^u_{ij}-\delta^d_{ij}, ~\cos \beta_u \equiv
c^u_\beta, ~~\cos \beta_d \equiv c^d_\beta.$

3.  We compare now the CKM matrix elements with experimental data
\cite{5}, first assuming  three generations and adjusting values
of  $\mu_i, m_i$ in  $\sa^u, \sbe^u, \sa^d, \sbe^d,$ and
$\Delta_{12}, \Delta_{13}$ to reproduce experimental bounds.

 For
the CKM matrix (13) one can simplify  replacing all $c^s_i,
i=\alpha, \beta,  ~~ s=u,d$ by unity (this amounts to $(1\div
2)$\% accuracy).

One takes values of $m_i$ at the scale 2 GeV  from \cite{5},
$m_u=2$ MeV, $m_d=5$ MeV, $m_c=1.25$ GeV, $m_s=95$ MeV, $m_b=4.54$
GeV, $m_t=173$ GeV.

>From (\ref{13}) one obtains \be \frac{\sa^u}{\sa^d} = \frac{
|V_{ub}|}{|V_{td}|} = 0.485 \pm 0.08,~~ \sa^d =\frac{
|V_{td}|}{|V_{ts}|} = 0.21 \pm 0.03,\label{14}\ee

$$\sa^u =\frac{
|V_{ub}|}{|V_{cb}|} = 0.095 \pm 0.012,$$ and the last two values
are compatible with the first one. Eqs. (\ref{14}), (\ref{12})
allow to derive $\mu_u\cong 13 $ MeV, $ \mu_d\cong 9 $ MeV.

Next is the case of $\sbe^u, \sbe^d$ which always enter in
(\ref{13}) with the relative phase $\Delta_{23}$,

\be |V_{cb}|\cong |V_{ts}|= |\sbe^d - e^{-i\Delta_{23}}\sbe^u| =0.0412\pm
0.0011.\label{15}\ee We assume at this point, that $\Delta_{23}\equiv 0$ (which
yields $\Delta_{12}=\Delta_{13})$,\footnote{This assumption implies that the CP
violating phase is generated by the interaction of lowest generation with the
nontrivial vacuum of the field $C_\mu$,and will be discussed at length
elsewhere.} and (\ref{15}) is satisfied using (\ref{12}) with $\mu_s=0.29$ GeV
and $\mu_c=12.25$ GeV. This yields $\sbe^u=0.252, \sbe^d=0.211. $ Finally,
$\mu_t, \mu_b$ are defined from the CMM condition (\ref{9}), $\mu_t\approx 160
$ GeV, $\mu_b \cong 4.2$ GeV.

One can now check all construction computing reparametrization invariant
quantities: angles $\alpha,\beta, \gamma,$ Wolfenstein parameters $\bar \rho,
\bar \eta$ \cite{6} and Jarlskog parameter $J$\cite{7}.

From(\ref{13}) $\alpha=\arg \left( -\frac{V_{tb}^*
V_{td}}{V_{ub}^*V_{ud}}\right) =\Delta_{12} =
\left(99^{+13}_{-8}\right)^\circ$\cite{5}, while \be  \beta= \arg\left(
-\frac{V_{cb}^* V_{cd}}{V_{tb}^*V_{td}}\right)=\arg (\sa^d -\sa^u
e^{-i\Delta_{12}}).\label{15a}\ee

Taking central values of $\sa^d, \sa^u$ from (\ref{14})  and $\alpha$ in the
experimental bounds, one obtains $0.67< \sin 2\beta <0.8$, which is compatible
with PDG value $\sin 2\beta =0.687 \pm 0.0324$. Another check is calculation of
$\bar \rho+i\bar \eta$ \cite{6} from (\ref{14}) and experimental value of
$\alpha =\Delta_{12}$,\be \bar \rho + i \bar \eta =-\frac{V_{ud}
V_{ub}^*}{V_{cd}V_{cb}^*}=\frac{\sa^u (\sa^u-\sa^d e^{-i\alpha})}{(\sa^u -\sa^d
e^{-i\alpha})^2} = 0.208+ i0.337.\label{16}\ee

This value is well compared with  experimental $(\bar \rho + i\bar\eta)_{exp} =
0.221^{+0.064}_{-0.028} + i 0.34^{+0.017}_{-0.045}$. Finally the Jarlskog
parameter for central values from (\ref{13}), (\ref{14}) is \be J= \sa^u \sa^d
|\sbe^d -\sbe^u e^{i\Delta_{23}}|^2\sin \Delta_{12} = 3.15\cdot
10^{-5}\label{17}\ee $vs$ experimental \cite{5} $J=
\left(3.08^{+0.16}_{-0.18}\right) \cdot 10^{-5}$.

We thus see, that CKM matrix (\ref{13}) provides a simple and
reasonable form in good agreement with  experimental data for
three generations.

The same is true also for the neutrino mixing matrix $V_{e\nu}$, and it is
demonstrated in Appendix 2, that $V_{e\nu}$ acquires the tribimaximal form,
when masses $\mu_i$ have reasonable values and $m_i$ satisfy experimental
bounds. The difference from the quark matrices lies in a very small value of
$s_\alpha^e \to 0$, ( while $s_\alpha^d\approx 0.2$), which leads to the
tribimaximal form of $V_{e\nu}$.

As it is, one can persuade oneself, that  the  last form of the CKM matrix in
(\ref{13}) is defined by three angles and one phase, e.g. by $s^4_\alpha,
s^d_\alpha$ and $\sin (\beta_u-\beta_d) \equiv  s^u_\beta c^d_\beta - s^d_\beta
c^u_\beta$,~~ and the phase $\Delta_{12}\equiv \alpha$. The three angle
parameters are expressed via $\mu_u, \mu_d$ and one combination of  $(\mu_c,
\mu_s$) (for $\sin (\beta_u -\beta_d))$. This corresponds to the actual number
of parameters in the $3\times 3$ unitary matrix, but   in our case all
parameters have a clear physical meaning.

 4. We turn now to the case of four generations.

First of all  one realizes, that the Pagels-Stokar relation \cite{8},\cite{1}
for our multiple stationary  solutions $\mu_i(p)$ of Eq. (\ref{5}) analyzed in
\cite{3}, has a generalized form

\be v^2 = \frac{N_c}{4 \pi^2} \sum^4_{i=1} \mu^2_i \ln
\frac{M}{\mu_i}, \label{18}\ee where $v=246$ GeV and
$\mu_i=\bar{\mu_i}(p)$ is the effective value of $\mu_i(p)$ in the
corresponding loop integral. Since one assumes that $M\gg\max
\mu_i= \mu_4$ the value of $\mu_4$ is limited from above and for
$M\geq 4 \mu_4$ one has $\mu_4 \lesssim 0.8$ TeV. Now one can
connect $\mu_i$ and $m_i$ in the same way, as it was done above
for three generations (see appendix 3 of \cite{3} for details).
One has \be m_4 = \frac{a_1}{2} + \sqrt{\frac{a^2_1}{4} - a_2}, ~~
m_3 = \frac{a_1}{2} - \sqrt{\frac{a^2_1}{4} - a_2}, a_1
=\sum^4_{i=1} \mu_i,\label{19}\ee
$$ a_2 \cong \mu_4 (\mu_2+\mu_3)\bar \eta + O(\eta, \delta).$$
here $\bar \eta =-\eta_{34}>0$, and if $\bar \eta\gg \frac14
\eta$, then relations (\ref{9}) for $m_1, m_2$ are approximately
the same, so that we keep the values $\mu_s, \mu_c, \mu_u,\mu_d$
of the previous section.

We take  two estimates for $\mu^u_3, \mu^u_4$: a) $\mu^u_4=0.8$ TeV,
$\mu^u_3=0.4$ TeV;\\ b) $\mu^u_4, \mu^u_3= (0.5, 0.3) $ TeV and obtain $\bar
\eta_4, s_4^u$ in two cases.

a) $\bar \eta_4=0.55, ~~ s^u_4=0.51; ~~ b) \bar \eta_u=0.7, ~~ s^u_4=0.54$. For
$\mu_3^d, \mu_4^d$ the situation is less constrained, since one can choose
$\mu_3^d\ll \mu^d_4$ and one has $s^d_4 \ll 1, \bar \eta_d \ll 1$
 and an equivalent limiting form
for the $ 4 \times 4$ matrix $W$ has three zeros: $W_{31} = W_{41} =W_{42} =0$

\be W=\left(
\begin{array}{llll}
c_\alpha,& -s_\alpha c_\beta e^{i\delta_{12}}, & c_4 s_\alpha
s_{\beta} e^{i\delta_{13}},& s_4s_\beta s_\alpha e^{i\delta_{14}}\\
-s_\alpha e^{-i\delta_{12}},& -c_\alpha c_\beta,& c_4 s_\beta
c_\alpha e^{i\delta_{23}},& s_4 s_\beta c_\alpha e^{i\delta_{24}}\\
0,& s_\beta e^{-i\delta_{23}},& c_4 c_\beta, & s_4c_\beta
e^{i\delta_{34}}\\
0,&0,& -s_4e^{-i\delta_{34}},&c_4\\\end{array}\right).\label{21}\ee Here the
phases $\delta_{ik}$ satisfy conditions

\be \delta_{24}=\delta_{23}+\delta_{34}, ~~ \delta_{14}
=\delta_{13}+\delta_{34}, ~~ \delta_{14}-\delta_{24}=\delta_{12}.\label{22}\ee

The $3\times 3$ matrix $W$ is easily obtained putting $s_4=0, c_4=1$.

Parameters $c_\alpha, c_\beta, c_4 (s_i=\sqrt{1-c^2_i}, i=\alpha, \beta, 4)$
are expressed via $m_i, \mu_i, i= 1,2,3,4$

\be c^2_\alpha =\frac{m_2-\mu_1}{m_2-m_1}, ~~ c^2_\beta
=\frac{m_3-\mu_2}{m_3+\mu_1- m_1-m_2},
~~c^2_4\cong\frac{\mu_4-m_3+\mu_2}{m_4-m_3+\mu_2}.\label{23}\ee

The resulting CKM matrix is given in Appendix 1.

One can check, whether CKM matrix is compatible with experiment in
the case of four generations. The sensitive points are unitarity
relations and the matrix element $V_{ud}$.

We list  as an example two  possible  sets of masses $\mu_i$ which define
together with $m_i$ the parameters $s_k^i, c_k^i, i=u,d; k=\alpha, \beta, 4$

$ \mu_d =10$ MeV, $\mu_s =0.29$ GeV, $\mu_u=13$ MeV, $\mu_c=12.25$ GeV; \\set
a) $\mu_t, \mu_{t'} = (400, 800)$ GeV; $\mu_b, \mu_{b'} = (20, 400$ GeV);
$m_{b'}, m_{t'}= (420, 1200) $ GeV;\\ set b) $\mu_t, \mu_{t'} = (300, 500)$
GeV; $\mu_b, \mu_{b'}=(200, 380)$ GeV; $m_{b'},  m_{t'} =(580, 630)$ GeV.

The resulting parameters are:

$s_\beta^u; c_\beta^u=0.252; 0.968; ~ s^u_\alpha; c_\alpha^u=0.095; 0.995$

$s_\beta^d; c_\beta^d =0.211; 0.98; ~ s_\alpha^d; c_\alpha^d=0.21; 0.98;$

 a)$s_4^u; c_4^u =0.51; 0.86;
~ s_4^d; c_4^d =  0.22,  0.97;$

b)$s_4^u; c_4^u =0.54; 0.85; ~ s_4^d; c_4^d =  0.57,  0.82.$

Let us first check, whether the $3\times 3$ part of the CKM matrix is not
deteriorated by the inclusion of the 4-th generation.

Taking $V_{ud}$ and parameter values from Appendix 1, one can write \be
V_{ud}=0.975+ 0.019\cdot  e^{i\Delta_{12}}+ 0.00084 e^{i\Delta_{13}} + 1.2
\cdot 10^{-4} e^{i\Delta_{14}}.\label{24}\ee

Assuming as before, that the nonzero phase is due to the first
generation only, i.e. $\Delta_{12} =\Delta_{13} = \Delta_{14} =
\alpha$ and $\alpha$ in experiment is close to $\frac{\pi}{2}$
\cite{5}, one obtains a good agreement with experimental value for
$V_{ud}, |V_{ud}| = 0.97418\pm 0.00027 $ \cite{5} with strong
limits on $\alpha$, $|\alpha-90^\circ|<1.5^\circ$.

Another CKM element known with good accuracy is $V_{us}$ and with the same
parameters from Appendix  1 one obtains for $\Delta_{12} = \alpha = 90^\circ$,
and the same assumptious about $\Delta_{ik} =0$ with $i,k\neq 1, | V_{us} | =
0.228$, which is close to the experimental value \cite{5}: $|V_{us}| =
0.2255\pm 0.0019$, and $|V_{cd}| =0.215~ vs$ experimental $|V_{cd}| = 0.230\pm
0.011$\cite{5}.
  In a similar way one can check other CKM coefficients which are
  known experimentally with larger errors and therefore this check
  is not  sensitive to the contribution of the fourth generation.

 5. At this point one can compare our results with that in the
  literature, for reviews and references see \cite{9,10}.

  First of all, we note, that our $4\times  4$ CKM matrix contains
  two additional  phases and two angles ($s^u_4, s^d_4)$ as
  compared to the $ 3\times 3
$ matrix. Moreover, in  the one-phase limit $(\Delta_{12} =\Delta_{13}
=\Delta_{14}, ~~ \Delta_{ik} =0, i,k\neq 1)$ no additional phase appears, and
$s^u_4, s^d_4$ and hence all $V_{ij}$ are defined by the masses $\mu^{u,d}_4,
m^{u,d}_4$, in addition to four parameters of $3\times 3$ matrix, and the
resulting scheme is rather rigid with six overall parameters, which is less
than 9 possible parameters of $4\times 4$ unitary matrix (with $2n-1$ phases
removed).

First of all, it is important to compare our results with the bounds
\cite{11}-\cite{14} on the mixing coefficients and masses $m_{b'}, m_{t'}$,
following from FCNC box diagrams and the decay $b\to s\gamma$, and also from
the so-called precise EW tests (EWPT). The  FCNC bound on mixing coefficients
$V_{ib'}, V_{t'i}$, were studied in \cite{14}, and can be written in our terms
(assuming as before the only nonzero CPV phase $\Delta_{1i}=\alpha$) as the
bound on the combination $K=-c^u_4s^d_4+s^u_4c^d_4< 0.6$ (for a ``conservative
bound'' \cite{14}). From our mixing coefficients above one has $K\cong 0.3$ for
the set a) and $K=0.042$ for the symmetric  set b). Thus both sets satisfy the
FCNC bounds of \cite{14}. However, as shown in \cite{13}, the EWPT bounds are
much more restrictive, and can be again reduced to the bound on the same
quantity $K(\cong s_{34}$ in notations of \cite{13}), e.g. $K\la 0.1$ (95\%
C.L.). One can see, that this limit is easily satisfied by our set b), where
masses $m_{b'}\approx m_{t'}\approx 0.6$ TeV and $m_{t'}-m_{b'}\approx 50$ GeV,
in the same range, as the values considered in \cite{12,14}. Thus one can see,
that the mixing $V_{tb'} \simeq K\approx -V_{t'b}\approx 0.1$ can be easily
accommodated in the  coherent mixing scheme, and it satisfies both FCNC and
EWPT bounds.
 On another
hand, as it is stressed in \cite{15}, this  mixing opens up an  interesting
possibility of the search for the 3-body decay of Higgs in processes like $H\to
\bar t' b W^+$ or $\bar b'tW^-$, where the wide composite Higgs is in the TeV
region.

Summarizing, a new mechanism producing several generations of quarks, and
mixing between them is suggested. The resulting CKM matrix in agreement with
all experimental data is found for three generations. In the case of four
generations stringent FCNC and EWPT bounds on masses and mixings drastically
limit available space  of mass  parameters. It is also interesting, that our
scheme provides a simple acceptable parametrization of the lepton mixing
matrix, as shown in Appendix 2.

As it was stressed recently in \cite{16}, in the general situation with
$m_{\nu'}\neq\frac12 m_{Z'}$, the case of four generations is still disfavored
in the global fit analysis, including oblique parameters $S, T,U$ and  FCNC
constraints. For possible implications of the latter see \cite{17}.

 Our scheme  predicts the EWSB connected primarily  with $b,t$ quarks in case of 3 and with the $b',t'$ quarks in case of four
generations (see e.g. Eq. (\ref{18})), and this is in common with approaches
derived in \cite{18}.

The author is grateful for useful discussion to V.A.Novikov, V.I.Shevchenko and
M.I.Vysotsky. The author acknowledges  financial support of the RFFI grant
09-02-00629a.

\vspace{2cm}
 \setcounter{equation}{0}
\renewcommand{\theequation}{A1.\arabic{equation}}

{\bf \large
\noindent Appendix  1}\\

\noindent{\bf \large The CKM $4\times 4$ matrix elements}\\

The CKM matrix $V=W_uW^+_d$ is obtained from (\ref{10}) with
parameters $\sa^u,\sbe^u, s_4^u, \sa^d, \sbe^d, s_4^d$  (and the
corresponding cosine terms), and phases $\delta^u_{ij},
\delta^d_{ij}$,  satisfying conditions (\ref{11}) and (\ref{22}).

$$V_{ud} = c^u_\alpha c^d_\alpha+\sa^u\sa^d
c^u_\beta c^d_\beta e^{i\Delta_{12}}+ c_4^u
 s^u_\alpha s^u_\beta c_4^ds^d_\beta  s^d_\alpha e^{i\Delta_{13}}+
 s_4^us_4^d\sbe^u\sa^u\sbe^d\sa^d e^{i\Delta_{14}};$$

$$ V_{us} = -c^u_\alpha s_\alpha^d   e^{i\delta_{12}^d} +\sa^u c^u_\beta
c_\alpha^d c^d_\beta   e^{i\delta^u_{12}}+
 c_4^u c_4^d s^u_\alpha s^u_\beta
\sbe^dc^d_\alpha  e^{i\delta^u_{13}-i\delta^d_{23}}+
s_4^us_4^d\sbe^u\sa^u\sbe^dc_\alpha^d
e^{i\delta^u_{14}-i\delta^d_{24}} $$

$$V_{ub}= -s^u_\alpha c^u_\beta\sa^d  e^{i\delta_{12}^u+\delta^d_{23}} +c_4^uc_4^d\sa^u
 s^u_\beta  c^d_\beta
e^{i\delta_{12}^u} + s_4^us_4^d \sbe^u\sa^uc_\beta^d
e^{i\delta^u_{14}- i\delta^d_{34}}$$
$$ V_{ub'}= - c_4^us_4^d\sa^u\sbe^u
e^{i\delta_{13}^u+i\delta_{34}^d}+ s_4^uc_4^d\sbe^u\sa^u
e^{i\delta^u_{14}}$$

\be V_{cd} =- s^u_\alpha  c^d_\alpha e^{-i\delta^u_{12}}+
c^u_\alpha c^u_\beta \sa^dc^d_\beta e^{-i\delta^d_{12}}+
c_4^uc_4^d\sbe^u c^u_\alpha\sa^d s^d_\beta
e^{i\delta^u_{23}-i\delta^d_{13}} +s_4^us_4^d\sbe^uc_\alpha^u
\sbe^d\sa^d e^{i\delta^u_{24}-i\delta^d_{14}} \label{A2.17}\ee

$$V_{cs}= s^u_\alpha   s^d_\alpha
 e^{-i\Delta_{12}}+ c_\alpha^u c_\beta^u c^d_\alpha
c^d_\beta + c_4^uc_4^d \sbe^uc^u_\alpha\sbe^d c^d_\alpha
e^{i\Delta_{23}}+s_4^us_4^d\sbe^uc^u_\alpha \sbe^d c_\alpha^d
e^{i\Delta_{24}}$$

$$V_{cb}=-c^u_\alpha c^u_\beta s^d_\beta e^{i\delta^d_{23}}
+ c_4^uc_4^d \sbe^u c^u_\alpha c^d_\beta e^{i\delta^u_{23}}+
s_4^us_4^d\sbe^uc_\alpha^uc_\beta^d e^{i\delta^u_{24}-i\delta^d_{34}} ,$$

$$ V_{cb'}
=-c_4^us_4^d\sbe^uc_\alpha^ue^{i\delta^u_{23}+i\delta^d_{34}}+s_4^uc_4^d\sbe^uc_\alpha^u
e^{i\delta^u_{24}}$$

$$V_{td}=-\sbe^u \sa^d c^d_\beta e^{-i\delta^u_{23}+i\delta^d_{12}} + c_4^uc_4^d    c^u_\beta  s^d_\alpha \sbe^d
e^{-i\delta_{13}^d} +s_4^us_4^d c^u_\beta s^d_\beta s^d_\alpha
e^{i\delta^u_{34}-i\delta^d_4},$$

$$V_{ts} = -  s^u_\beta c^d_\alpha
c^d_\beta e^{-i\delta_{23}^u} +  c_4^uc_4^d  c^u_\beta\sbe^d c^d_\alpha
e^{-i\delta^d_{23}} + s_4^us_4^d c^u_\beta \sbe^d c_\alpha^d
e^{i\delta^u_{34}-i\delta^d_{24}},$$

$$V_{tb} =
s^u_\beta s^d_\beta e^{-i\Delta_{23}}+c_4^uc_4^d  c^u_\beta
c^d_\beta+s_4^us_4^d c_\beta^u c_\beta^d e^{i\Delta_{34}} ,$$

$$ V_{tb'} = - c_4^u s_4^d c_\beta^u e^{i\delta^d_{34}}+ s_4^u
c_4^d c_\beta^u e^{i\delta^u_{34}}$$

$$ V_{t'd} =- s_4^u c_4^d \sa^d \sbe^d
e^{-i\delta^u_{34}-i\delta^d_{13}}+ c_4^u s_4^d \sbe^d \sa^d
e^{-i\delta^d_{14}}$$

$$ V_{t's} =- s_4^u c_4^d \sbe^dc_\alpha^d
e^{-i\delta^u_{34}-i\delta^d_{23}}+ c_4^u s_4^d \sbe^d c_\alpha^d
e^{-i\delta^d_{24}}$$

$$ V_{t'b} =- s_4^u c_4^d c_\beta^d
e^{-i\delta^u_{34}}+ c_4^u s_4^d c_\beta^d e^{-i\delta^d_{34}}$$

$$ V_{t'b'} =  s_4^u s_4^d
e^{-i\Delta_{34}}+ c_4^u c_4^d $$

Parameter values $\sa^u,\sa^d, \sbe^u,\sbe^d$ were discussed in the text, Eq.
(\ref{23}) and below.

 \vspace{2cm}
 \setcounter{equation}{0}
\renewcommand{\theequation}{A2.\arabic{equation}}

{\bf \large
\noindent Appendix 2 }\\

\noindent{\bf \large Coherent mixing in leptons}\\

In this appendix we apply the CMM matrices to families of $e$- and
$\nu$- sectors. As in the main text and in \cite{3} we define the
$\hat W_e, \hat W_\nu$ matrices in the following way \be \hat
\mu_{e,\nu} = \hat W^+_{e,\nu} m_{e,\nu} (diag) \hat
W_{e,\nu}\label{A2.1}\ee so that the mixing matrix is $\hat
V_{e,\nu}= \hat W_e \hat W_\nu^+$.

In the  $e$-sector with masses $m_e =0.51$ MeV, $m_\mu =105.66$
MeV and $m_\tau =1777 $ MeV.

Choosing $\mu_e\cong 1$ MeV, and not fixing yet  $\mu_\mu$  and $\mu_\tau$, one
obtains the matrix $\hat W_e$ in the simplified from (\ref{10}), where we put
$c_\alpha\to c_\alpha^e, ~~ s_\alpha\to s^e_\alpha, ~~ c_\beta \to c_\beta^e,
~~ s_\beta \to - s_\beta^e.$ We shall neglect for simplicity  possible Majorana
phases, and obtain using (\ref{12}) \be s^e_\alpha \cong 0.07 \to 0 ,~~
c_\alpha^e \cong 1; ~~ s^e_\beta , ~~ c^e_\beta .\label{A2.2}\ee In the same
way we are obtaining the matrix $\hat W_\nu$ as in (\ref{12}) replacing
$s_\beta$ in (\ref{10}) by ($+ s^\nu_\beta)$, and we have \be  V_{e\nu}
(s^e_\alpha\to 0)=\left(
\begin{array}{lll}
c_\alpha^\nu& s_\alpha^\nu e^{i\delta^\nu_{12}}&0\\
-s_\alpha^\nu \cos(\beta_e +\beta_\nu)e^{-i\delta^\nu_{12}}&
c^\nu_\alpha \cos (\beta_e+\beta_\nu)& -\sin (\beta_e
+\beta_\nu)\\

-s^\nu_\alpha \sin (\beta_e+\beta_\nu)e^{-i\delta^\nu_{13}}& c^\nu_\alpha\sin
(\beta_e+\beta_\nu) &\cos (\beta_e+\beta_\nu)
\end{array}\right),\label{A2.3}\ee
where $\cos \beta_e\equiv c_\beta^e,~~ \cos \beta_\nu\equiv
c_\beta^\nu,$ and using experimental data \cite{5}, $\sin
\theta_{13} <5.10^{-2},~~ \sin^2( 2 \theta_{23})>0.90, ~~ tg^2
\theta_{12}=0.47^{+0.06}_{-0.05}$, one has

 \be s_\alpha^\nu
\cong \frac{1}{\sqrt{3}}, ~~ c_\alpha^\nu\cong\sqrt{\frac{2}{3}},
~~ \beta_e+\beta_\nu \cong \frac{\pi}{4} \label{A2.4}\ee  In this
way one reproduces  the well-known tribimaximal matrix
$V_{e\nu}(s_\alpha^e \to 0) = U_{TB},$ where \be U_{TB}
=\left(\begin{array}{ccc} \sqrt{\frac23}& \frac{1}{\sqrt{3}}&0\\
-\frac{1}{\sqrt{6}}&
\frac{1}{\sqrt{3}}&-\frac{1}{\sqrt{2}}\\
-\frac{1}{\sqrt{6}}&\frac{1}{\sqrt{3}}&\frac{1}{\sqrt{2}}\end{array}\right)\label{A2.4a}\ee

At this point one should find $\mu_\mu, \mu_\tau$ and the set $\mu^\nu_i, ~~
i=1,2,3,$  which correspond to the conditions (\ref{A2.4}). Having in mind,
that $s_\beta^e = \sqrt{\frac{{m_\tau -\mu_\tau}}{2 m_\tau-\mu_\tau -\mu_\mu}}
\approx \sqrt{ \frac{\mu_\mu}{m_\tau}}\geq \sqrt{\frac{m_\mu}{m_\tau}} \approx
0.25$ and $s^\nu_\beta \geq \sqrt{\frac{m_2}{m_3}} \approx 0.42$, (for
experimental values \cite{5} $m_2 \approx 0.875\cdot 10^{-2}$ eV, ~~ $m_3
\approx 4.9\cdot 10^{-2}$ eV, and assuming $ m_{2,3}\gg m_1$) one has a narrow
interval, $\beta_e \ga 15^\circ, ~~ \beta_\nu \ga 25^\circ, ~~
\beta_e+\beta_\nu=45^\circ$, and $\mu_3^\nu$ varies in the interval $0.75 m_3
\leq \mu^\nu_3\leq 0.82 m_3$ ;~~$m_\mu \leq \mu_\mu\leq2 m_\mu$. For the first
generation of $\nu$ both $\mu^\nu_1 >m_1$ should only be much smaller, than
$m_2, m_3$.

 For more discussion with the use of $W$ matrices of the type of
 Eq. (\ref{10}) see \cite{4}.

\end{document}